\documentclass[12pt,eqsecnum]{article}
\usepackage[dvips]{graphicx}
\usepackage{amssymb}
\usepackage{amsmath}
\usepackage{epstopdf}
\makeatletter

\@addtoreset{equation}{section}
\makeatletter

\newcommand{\ma}[1]{\mbox{$\mathcal{#1}$}}

\newcommand{\dalm}{\kern1pt\vbox{\hrule height 0.9pt\hbox{\vrule width
0.9pt\hskip 2.5pt\vbox{\vskip 5.5pt}\hskip 3pt\vrule width 0.3pt}\hrule height
0.3pt}\kern1pt}

\def\b2hat{ {\hat b}_2 }

\begin{document}

\begin{titlepage}
\vfill
\begin{flushright}
\today
\end{flushright}

\vfill
\begin{center}
\baselineskip=16pt
{\Large\bf 
Lifshitz black holes in Brans-Dicke theory
}
\vskip 0.5cm
{\large {\sl }}
\vskip 10.mm
{\bf Hideki Maeda${}^{a}$ and Gaston Giribet${}^{b}$} \\

\vskip 1cm
{
	${}^a$ Centro de Estudios Cient\'{\i}ficos (CECs), Casilla 1469, Valdivia, Chile \\
	${}^b$ Instituto de F\'{\i}sica de Buenos Aires, CONICET, Buenos Aires, Argentina.\\
	\texttt{hideki-at-cecs.cl, gaston-at-df.uba.ar}

     }
\vspace{6pt}
\end{center}
\vskip 0.2in
\par
\begin{center}
{\bf Abstract}
 \end{center}
\begin{quote}
We present an exact asymptotically Lifshitz black hole solution in Brans-Dicke theory of gravity in arbitrary $n(\ge 3)$ dimensions in presence of a power-law potential.
In this solution, the dynamical exponent $z$ is determined in terms of the Brans-Dicke parameter $\omega$ and $n$.
Asymptotic Lifshitz condition at infinity requires $z>1$, which corresponds to $-(n-1)/(n-2) \le \omega < -n/(n-1)$.
On the other hand, the no-ghost condition for the scalar field in the Einstein frame requires $0<z \le 2(n-2)/(n-3)$.
We compute the Hawking temperature of the black hole solution and discuss the problems encountered and the proposals in defining its thermodynamic properties.
A generalized solution charged under the Maxwell field is also presented.
  \vfill
\vskip 2.mm
\end{quote}
\end{titlepage}




\tableofcontents

\section{Introduction}
One of the most interesting ideas that have been proposed in the last years is that of extending the AdS/CFT holographic duality \cite{Malda, Witten, GKP} to the area of condensed matter physics. This idea has attracted considerable attention, and much 
effort has been made to construct holographic gravity duals for such systems
\cite{Kachru,BalasubramanianMcGreevy,BalasubramanianMcGreevy3,Son}. The reason why this area has became one of the most active ones in the field is that it raises the hope to explicitly work out the details of the non-perturbative dynamics of interesting scale invariant non-relativistic models. Among the class of models one could try to study with this holography inspired methodology, there are systems of strongly correlated electrons, which have received a renewed interest recently and permit to address questions that are accessible experimentally.

In particular, for the case of systems exhibiting anisotropic scale invariance, but not Galilean invariance, there exists a concise proposal for their holographic counterparts. Such models, known as Lifshitz fixed points, present symmetry under 
\begin{equation}
t \to L^{2z} t \ , \ \ \ x^i \to L^{2} x^i.     \label{Gscaling}
\end{equation}
where $L$ is the scale parameter and $x^i$ are the spatial coordinates $i=1,2,... d$; here we will use the convention $d=n-2$. The time and the spatial coordinates suffer from different scaling transformations, and this is characterized by the value of the dynamical exponent $z$. 

Among the physical systems that exhibit symmetry under (\ref{Gscaling}) one finds several strongly correlated models in condensed matter, like the Rokhsar-Kivelson dimer model in $d=2$ dimensions. Critical points of this kind were also conjectured to exist in $d=1$; such is the case of the transition from a Luttinger liquid to a ferromagnetic state. Being strongly coupled models, having a tool to compute observables, such as correlation functions, is clearly of importance. For instance, the question as to whether the previously reported ultralocal behavior of correlation functions in $z=2$ models manifest itself or not could be addressed if a holographic calculation of these observables was at hand. 

In \cite{Kachru}, the spacetime candidates to represent gravity duals for this type of ($n-1$)-dimensional scale invariant fixed points were proposed. The metrics of these $n$-dimensional spacetimes naturally depend on the dynamical exponent $z$ of the model they want to be in correspondence with. The metric takes a simple form 
\begin{equation}
ds^2 = - \frac{r^{2z}}{l^{2z}}  dt^2 + \frac{l^{2}}{r^{2}}  dr^2 + r^2 dx^i dx_i,    \label{Gmetric}
\end{equation}
where $x^i$ are coordinates of the $d=n-2$ flat directions, with $i=1,2,...n-2$, and where $l$ and $z$ are real parameters. Metric (\ref{Gmetric}) generalizes that of $n$-dimensional Anti-de Sitter spacetime (AdS$_{n}$), which corresponds to the special case $z=1$. 

Then, (\ref{Gmetric}) happens to realize the symmetry transformation (\ref{Gscaling}) geometrically, provided a similar transformation on the bulk coordinate $r$ is performed as well; namely $r \to L^{-2} r$. In the holographic realization, the dual scale invariant theory would be formulated on a $n-1$-dimensional space that can be regarded as being located at infinite $r$, say the boundary, whose physical coordinates are the time $t$ and the $n-2$ spatial directions $x^i$. This is totally analog to AdS/CFT, and it happens to coincide with it when $z=1$. In fact, the holographic recipe follows the lines of the standard AdS/CFT prescription of \cite{Witten,GKP}. In particular, one can compute the scaling dimensions of the dual fields in the boundary theory by studying the asymptotic dynamics of fields in the bulk. For example, if one considers a scalar field $\varphi $ of mass $m$ in the bulk, and one demands suitable boundary conditions obeying a specific falling-off
\begin{equation}
    \varphi(t,x,r) \sim  \varphi_0(t,x) \ r^{-\Delta} + {\mathcal O} (r^{-\Delta-1}) , \label{Gasymptotic}
\end{equation}
it is easy to derive the scaling dimension $\Delta $ of the field $O_{\varphi }$ dual to $\varphi $, which turns out to be
\begin{equation}
\Delta _{\pm } = \frac{1}{2} ( (z+n-2) \pm \sqrt{(z+n-2)^2 + 4m^2 l^2 } \ ) . \label{GDelta}
\end{equation}

In the original paper \cite{Kachru}, the two-point function of operators $O_{\varphi }$ was calculated in the case of $d=2$ Lifshitz point with $z=2$. This function was shown to have the expected functional dependence 
\begin{equation}
\langle O_{\varphi }(x_1,t_1) \ O_{\varphi }(x_2,t_1) \rangle \ \sim | x_1 - x_2 |^{-8}.    \label{GKachru}
\end{equation}
Notice that for $n=4$ and $z=2$ (\ref{GDelta}) actually gives $\Delta _+ = 4$.

In \cite{BalasubramanianMcGreevy2}, on the other hand, the system at finite temperature $T$ was studied as well, and it was shown there that, in the regime of large separation, the expected exponential damping induced by the finite temperature effects arises; namely
\begin{equation}
\langle O_{\varphi }(x_1,t_1) \ O_{\varphi }(x_2,t_1) \rangle \ \sim  e^{-\sqrt{4\pi T} | x_1 - x_2 |} \ | x_1 - x_2 |^{-3/2} ; 
\end{equation}
while in the short distance regime the limit (\ref{GKachru}) is consistently recovered, and no evidence of ultralocality was found.

Also resembling the standard AdS/CFT correspondence, adding finite temperature to the problem amounts to replace the Lifshitz metric (\ref{Gmetric}) by a black hole solution that asymptotes to it. Then, the Hawking temperature $T$ of such a Lifshitz black hole would correspond to the temperature of the boundary scale invariant dual model. Nevertheless, an immediate obstruction appears here: Because of the form of the metric (\ref{Gmetric}) and the needs of holographic realization, such an asymptotically Lifshitz black hole would be a static spherically symmetric solution with a $z$-dependent asymptotic, and, usually, Birkhoff-like theorems and no-hair theorems prevent such spaces from appearing in natural models. In fact, the Lifshitz black holes found so far use to exist in models whose matter contents correspond to contrived stress-energy tensors. A manifestation of this is how specific the fine tuning of the coupling constants in the models analyzed in the literature has to be for Lifshitz black holes to be admitted as exact solutions. 

Lifshitz black holes were extensively discussed in the literature in the last two years, and both analytic and numerical solutions were found 
\cite{Danielsson, Danielsson2, Bertoldi, Takayanagi, BalasubramanianMcGreevy2}. Models involving Einstein-Hilbert action coupled to matter were 
analyzed, together with models consisting of higher-curvature modifications \cite{Ootro1, Gaston, Tonni, R2, Gaston2, Dehghani, Dehghani2, Marika, Wissam, Ootro1, Ootro2, refereequiere1, refereequiere2}. Charged Lifshitz 
black holes were also reported, and their thermodynamics and consistency studied \cite{Dehghani3, raros, Bertoldi3, Bertoldi4, Bertoldi5,
Pang, consistency}.  

In this paper, we will be concerned with Lifshitz black holes in Brans-Dicke theory. 
Brans-Dicke theory is the prototype of all the scalar-tensor theories of gravity and parametrized by the so-called Brans-Dicke parameter $\omega$~\cite{bd}.
This theory was originally formulated in four dimensions and the theory with sufficiently large $\omega$ passes all the observational and experimental tests for gravitation theories~\cite{will}.
(See~\cite{faraoni1999} for the relation between Brans-Dicke theory and general relativity.)
This theory is also interesting theoretically because it contains the dilatonic theory ($\omega=-1$) or $f(R)$ gravity ($\omega=0$ {in the presence of a potential})~\cite{fujiimaeda}.
In the asymptotically flat case in vacuum, Hawking showed a ``no-scalar-hair theorem'' claiming that a black hole cannot support a non-trivial Brans-Dicke scalar field if the scalar field is regular on and outside the event horizon~\cite{hawking1972}.
Therefore, in order to have a hairy black hole configuration in Brans-Dicke theory, one has to consider (I) non-asymptotically flat spacetime, (II) non-regular scalar field on the horizon, (III) potential for the scalar field, or (IV) other matter fields~\cite{dph2006,bdbh}.

We will show that asymptotically Lifshitz solutions exist in this theory provided the scalar field potential to have a 
specific, though simple, functional form. The motivation is to show that analytic Lifshitz black hole solutions with a 
continuous and finite range of $z$, and in arbitrary number of dimensions $n$ do exist in a familiar field theory, 
where the value of $z$ is determined by $\omega$ and $n$. Here, we report black holes with or without a Maxwell field, 
whose metrics can be regarded as generalizations of solutions previously reported in the literature.

The paper is organized as follows.
In section II, we present our assumptions and derive the basic equations.
In section III, we derive an exact asymptotically Lifshitz vacuum solution and discuss its properties.
In section IV, we generalize the solution in the presence of a Maxwell field.
Our conclusions and future prospects are summarized in section IV.
In appendix A, we present a vacuum solution which is a generalization of the Brans solution in the four-dimensional spherically symmetric spacetime and show that it certainly contains the (asymptotically) Lifshitz vacuum solution. 
In appendix B, we show that there is no Lifshitz solution in the Einstein frame with a real scalar field.
{In appendix C, the solution for the AdS case ($z=1$) is presented.}
Our basic notations follow \cite{wald}.
The conventions of curvature tensors are 
$[\nabla _\rho ,\nabla_\sigma]V^\mu ={R^\mu }_{\nu\rho\sigma}V^\nu$ 
and $R_{\mu \nu }={R^\rho }_{\mu \rho \nu }$.
The Minkowski metric is taken to be the mostly plus sign, and 
Roman indices run over all spacetime indices.

\section{Preliminaries}
\subsection{Brans-Dicke theory with a Maxwell field}
In this paper, we consider Brans-Dicke theory with a Maxwell field in arbitrary $n(\ge 3)$ dimensions.
We include a potential $V(\phi)$ for a Brans-Dicke scalar field $\phi$ in the action, which is given by 
\begin{align}
I=&\int d^nx\sqrt{-g}\biggl[\frac{1}{16\pi}\biggl(\phi R-\omega \frac{(\nabla\phi)^2}{\phi}-V(\phi)\biggl)-\frac{1}{\zeta^2}F_{\mu\nu}F^{\mu\nu}\biggl],\label{action}
\end{align}
where $(\nabla\phi)^2:=\nabla_\mu\phi\nabla^\mu\phi$, $\omega$ is the Brans-Dicke parameter, and $\zeta$ is a real coupling constant.
$F_{\mu\nu}:=\partial_\mu A_\nu-\partial_\nu A_\mu$ is the Faraday tensor defined by the gauge potential $A^\mu$, 
The field equations given from this action are
\begin{align}
G_{\mu\nu}=&\frac{8\pi}{\phi}(T^{(\phi)}_{\mu\nu}+T^{({\rm em})}_{\mu\nu}),\label{beq1} \\
\nabla_\nu F^{\mu\nu}=&0,\label{max}\\
\dalm\phi=&\frac{1}{2(n-2)\omega+2(n-1)}\biggl((n-2)\phi\frac{dV}{d\phi}-nV-\frac{16\pi(n-4)}{\zeta^2}F_{\rho\sigma}F^{\rho\sigma}\biggl),\label{beq2}
\end{align}
where $\dalm\phi:=\nabla_\mu\nabla^\mu\phi$, $G_{\mu\nu}:=R_{\mu\nu}-(1/2)g_{\mu\nu}R$ is the Einstein tensor, and 
\begin{align}
T^{(\phi)}_{\mu\nu}:=&\frac{1}{8\pi}\biggl[\frac{\omega}{\phi}\biggl(\nabla_\mu \phi\nabla_\nu \phi-\frac12g_{\mu\nu}(\nabla\phi)^2\biggl)-\frac12Vg_{\mu\nu}+(\nabla_\mu \nabla_\nu \phi-g_{\mu\nu}\dalm \phi)\biggl], \\
T^{({\rm em})}_{\mu\nu}:=&\frac{4}{\zeta^2}\biggl(F_{\mu}^{~\rho}F_{\nu\rho}-\frac14g_{\mu\nu}F_{\rho\sigma}F^{\rho\sigma}\biggl).
\end{align}
The trace of the gravitational equation (\ref{beq1}) gives
\begin{align}
R=&\omega \frac{(\nabla\phi)^2}{\phi^2}-\frac{n}{(2-n)\phi}V+\frac{2(1-n)}{2-n}\frac{\dalm\phi}{\phi}+\frac{16\pi(4-n)}{(2-n)\zeta^2\phi}F_{\mu\nu}F^{\mu\nu}.\label{trace}
\end{align}

As seen in the action (\ref{action}), the inverse of $\phi$ is the effective gravitational constant and therefore we assume that $\phi$ is positive.
The action with $\omega=-1$ gives dilatonic gravity shown as follows.
By defining $\Phi:=-\ln\phi$, we obtain the gravitational part of the action (\ref{action}) with $\omega=-1$ in the form of
\begin{align}
I=&\frac{1}{16\pi}\int d^nx\sqrt{-g}e^{-\Phi}\biggl(R+(\nabla\Phi)^2-W(\Phi)\biggl),\label{dilaton}\\
W(\Phi):=&e^{\Phi}V(\phi).
\end{align}
This action with $W(\Phi)=0$ and $n=10$ appears as the dilaton-graviton sector in the low-energy action of string theories~\cite{lowenergy}.

The action (\ref{action}) for $(g_{\mu\nu},\phi)$, which is called the Jordan frame, can be conformally transformed into the Einstein frame for $({\hat g}_{\mu\nu},\psi)$, where $\psi$ is a minimally coupled scalar field.
By the conformal transformation $g_{\mu\nu}=\Omega(\psi)^2{\hat g}_{\mu\nu}$ with 
\begin{align}
\Omega(\psi)=&(G_n\phi)^{1/(2-n)},\\
\phi=&\exp\biggl(\pm \sqrt{\frac{8\pi G_n(n-2)}{(n-1)+(n-2)\omega}}(\psi-\psi_0)\biggl),
\end{align}
where $G_n$ and $\psi_0$ are constants, the gravitational action in the Jordan frame (\ref{action}) is transformed into the Einstein frame as
\begin{align}
{\hat I}=&\int d^nx\sqrt{-{{\hat g}}}\left[\frac{1}{16\pi G_n}{{\hat R}}-\frac12 ({\hat {\nabla}}\psi)^2-{\hat V}(\psi)\right],
\end{align}
where $\hat {\nabla}$ is covariant derivative in the Einstein frame and the potential in the Einstein frame is given by
\begin{align}
{\hat V}(\psi)=&\frac{1}{16\pi}\Omega^{n}{V}(\phi),\nonumber \\
=&\frac{1}{16\pi}(G_n\phi)^{n/(2-n)}{V}(\phi).
\end{align}
It is seen that $G_n$ is the $n$-dimensional gravitational constant in the Einstein frame.
If the Brans-Dicke scalar field is real, $\psi$ is real only if 
\begin{align}
\omega \ge -\frac{n-1}{n-2}.\label{omega-eq}
\end{align}
$\psi$ becomes a ghost scalar field for $\omega<-(n-1)/(n-2)$.
In the present paper, we impose the condition (\ref{omega-eq}) for the physical theory.

\subsection{Ans{a}tze}
In this paper we consider an $n$-dimensional spacetime $({\ma M}^n, g_{\mu \nu })$ which is a
warped product of an $(n-2)$-dimensional maximally symmetric space $(K^{n-2},\gamma _{ij})$ and a two-dimensional orbit spacetime $(M^2, g_{ab})$ under the isometries of $(K^{n-2}, \gamma _{ij})$. 
The line element in this spacetime may be written as
\begin{align}
g_{\mu \nu }d x^\mu d x^\nu =g_{ab}(y)d y^ad y^b +r^2(y) \gamma
_{ij}(z) d z^id z^j , \label{eq:ansatz}
\end{align}
where $a,b = 0, 1$ while $i,j = 2, ..., n-1$. Here $r$ is a scalar
on $(M^2, g_{ab})$ and
$\gamma_{ij}$ is the metric on $(K^{n-2}, \gamma _{ij})$ with
its sectional curvature $k = \pm 1, 0$.

The $(n-2)$-dimensional maximally symmetric space satisfies
\begin{eqnarray}
\overset{(n-2)}{R}{}_{ijkl}=k(\gamma_{ik}\gamma_{jl}-\gamma_{il}\gamma_{jk}).
\end{eqnarray}
The superscript $(n-2)$ means that the geometrical quantity are defined
on $(K^{n-2}, \gamma _{ij})$. 
The non-zero components of the Einstein tensor are given by 
\begin{align}
G_{ab} = & -(n-2)\frac{D_aD_br-(D^2r)g_{ab}}{r}- \frac{(n-2)(n-3)}{2}g_{ab}\left(  \frac{k- (Dr)^2}{r^2} \right),\label{Gab} \\
G^{i}_{~j}=&-\frac{1}{2}\biggl[(n-3)(n-4) \left(  \frac{k- (Dr)^2}{r^2} \right)-2(n-3) \frac{(D^2 r)}{r}+\overset{(2)}{R}{}\biggl]\delta^i_{~j},\label{Gij}
\end{align}
Here $D_a$ is a metric compatible linear connection on $(M^2, g_{ab})$, $(Dr)^2:=g^{ab}(D_ar)(D_br)$, and $D^2r:=D^aD_ar$.
The contraction was taken over on the two-dimensional orbit space and ${}^{(2)}{R}$ is the Ricci scalar on $(M^2, g_{ab})$.

In this symmetric spacetime, $\phi$ is a scalar on $(M^2, g_{ab})$.
We obtain the non-zero components of $\nabla^\mu\nabla_\nu\phi$ as
\begin{align}
\nabla^a\nabla_b\phi=D^aD_b\phi,\qquad \nabla^i\nabla_j\phi=\frac{D_ar}{r}D^a\phi\delta^i_{~j}
\end{align}
and then
\begin{align}
\dalm\phi=D^2\phi+(n-2)\frac{D_ar}{r}D^a\phi.
\end{align}
The non-zero components of ${T^{(\phi)}}^{\mu}_{~\nu}$ are given by
\begin{align}
{T^{(\phi)}}^{a}_{~b}=&\frac{1}{8\pi}\biggl[\frac{\omega}{\phi}\biggl(D^a\phi D_b \phi-\frac12\delta^a_{~b}(D\phi)^2\biggl)-\frac12V\delta^a_{~b}+D^aD_b\phi-\delta^a_{~b}\biggl(D^2\phi+(n-2)\frac{D_dr}{r}D^d\phi\biggl)\biggl], \\
{T^{(\phi)}}^{i}_{~j}=&\frac{1}{8\pi}\biggl[-\frac{\omega}{2\phi}(D\phi)^2-\frac12V-D^2\phi-(n-3)\frac{D_dr}{r}D^d\phi\biggl]\delta^{i}_{~j}.
\end{align}

We assume that the electromagnetic field has the form
\begin{align}
A_{\mu}dx^\mu=A_a(y)dy^a+A_i(z)dz^i,
\end{align}
which in turn implies that the Faraday tensor reads
\begin{align}
F_{\mu \nu }d x^\mu \wedge d x^\nu = F_{ab}(y)d y^a \wedge d
y^b+F_{ij}(z)d z^i \wedge d z^j.
\end{align}
Here $F_{ab}(y)$ and $F_{ij}(z)$ are identified with the electric
and magnetic components, respectively. 
For the magnetic component, the compatibility with the field equations requires the following form;
\begin{align}
\gamma^{kl}F_{ik}F_{jl}=Q_{\rm m}^2\gamma_{ij},\label{mag-F}
\end{align}
where $Q_{\rm m}$ is a real constant~\cite{opz2008,mhm2010}. 
The Maxwell invariant scalar is given as
\begin{align}
F_{\mu\nu}F^{\mu\nu}=F_{ab}F^{ab}+\frac{(n-2)Q_{\rm m}^2}{r^4},
\end{align}
while the non-zero components of ${T^{({\rm em})}}^{\mu}_{~\nu}$ are given by
\begin{align}
{T^{({\rm em})}}^{a}_{~b}=&\frac{1}{\zeta^2}\biggl(F_{cd}F^{cd}-\frac{(n-2)Q_{\rm m}^2}{r^4}\biggl)\delta^a_{~~b}, \\
{T^{({\rm em})}}^{i}_{~j}=&-\frac{1}{\zeta^2}\biggl(F_{cd}F^{cd}+\frac{(n-6)Q_{\rm m}^2}{r^4}\biggl)\delta^i_{~~j},
\end{align}
where we used $F_{cd}F^{cd}=2F_{01}F^{01}$.

\subsection{Decomposed field equations}
Now we are ready to derive the decomposed field equations.
The gravitational equation (\ref{beq1}) gives
\begin{align}
&-(n-2)\frac{D^aD_br-(D^2r)\delta^a_{~b}}{r}- \frac{(n-2)(n-3)}{2}\delta^a_{~b}\left(  \frac{k- (Dr)^2}{r^2} \right) \nonumber \\
&~~~~~~=\frac{1}{\phi}\biggl[\frac{\omega}{\phi}\biggl(D^a\phi D_b \phi-\frac12\delta^a_{~b}(D\phi)^2\biggl)-\frac12V\delta^a_{~b}+D^aD_b\phi-\delta^a_{~b}\biggl(D^2\phi+(n-2)\frac{D_dr}{r}D^d\phi\biggl)\biggl] \nonumber \\
&~~~~~~~+\frac{8\pi}{\zeta^2\phi}\biggl(F_{cd}F^{cd}-\frac{(n-2)Q_{\rm m}^2}{r^4}\biggl)\delta^a_{~~b},\label{eq1}\\
&-\frac{1}{2}\biggl[(n-3)(n-4) \left(  \frac{k- (Dr)^2}{r^2} \right)-2(n-3) \frac{(D^2 r)}{r}+\overset{(2)}{R}{}\biggl] \nonumber \\
&~~~~~~=\frac{1}{\phi}\biggl[-\frac{\omega}{2\phi}(D\phi)^2-\frac12V-D^2\phi-(n-3)\frac{D_dr}{r}D^d\phi\biggl]-\frac{8\pi}{\zeta^2\phi}\biggl(F_{cd}F^{cd}+\frac{(n-6)Q_{\rm m}^2}{r^4}\biggl),\label{eq2}
\end{align}
while the scalar field equation (\ref{beq2}) reduces to
\begin{align}
D^2\phi+(n-2)\frac{D_ar}{r}D^a\phi=&\frac{1}{2(n-2)\omega+2(n-1)}\biggl[(n-2)\phi\frac{dV}{d\phi}-nV \nonumber \\
&-\frac{16\pi(n-4)}{\zeta^2}\biggl(F_{ab}F^{ab}+\frac{(n-2)Q_{\rm m}^2}{r^4}\biggl)\biggl].\label{eq3}
\end{align}
The Maxwell equation (\ref{max}) reduces to
\begin{align}
D_bF^{ab}=0,\qquad {\bar D}_jF^{ij}=0,\label{max-d}
\end{align}
where ${\bar D}_j$ is a metric compatible linear connection on $(K^{n-2}, \gamma _{ij})$. 
By eliminating the potential $V$ from Eqs.~(\ref{eq1}) and (\ref{eq2}), we obtain the following useful equation without the potential term; 
\begin{align}
&-(n-4)\frac{(D^2r)}{r}-2(n-3)\left(  \frac{k- (Dr)^2}{r^2} \right)+\overset{(2)}{R}{} \nonumber \\
&~~~~~~=\frac{1}{\phi}\biggl[\frac{\omega}{\phi}(D\phi)^2+D^2\phi-2\frac{D_dr}{r}D^d\phi\biggl]+\frac{32\pi}{\zeta^2\phi}\biggl(F_{ab}F^{ab}-\frac{2Q_{\rm m}^2}{r^4}\biggl).\label{eq4}
\end{align}
The decomposed field equations (\ref{eq1})--(\ref{eq4}) are the basic equations in this study.

\section{Exact asymptotically Lifshitz vacuum black hole solutions}
In this section, we consider the vacuum case, namely the case with $T^{({\rm em})}_{\mu\nu}\equiv 0$.
Hereafter, we focus our attention on the case with $k=0$ because it allows for Lifshitz solution.

\subsection{Causal structure of the Lifshitz spacetime}
{To begin, let us clarify the basic properties of the background Lifshitz spacetime.}
The Lifshitz spacetime is defined by 
\begin{align}
ds^2=-\frac{r^{2z}}{l^{2z}}dt^2+\frac{l^2}{r^2}dr^2+r^2dx^idx_i,
\label{lifshitz-n}
\end{align} 
where $z$ and $l$ are real constants {and $l$ is assumed to be positive}.
$dx^idx_i$ is the line element of the $(n-2)$-dimensional flat space.
{We note that the constant $l$ cannot be set to one by rescaling of the coordinates.}
{The curvature invariants are all constant in this spacetime, however, some of the components of the Riemann tensor in the parallelly propagated orthonormal frame blows up at $r=0$ unless $z=1$~\cite{cm2011}.
Namely, there is a curvature singularity at $r=0$ if $z\ne 1$.}

{The non-zero components of the Einstein tensor is 
\begin{align}
G^t_{~t}=&\frac{(n-1)(n-2)}{2l^2},\\
G^r_{~r}=&\frac{(n-3+2z)(n-2)}{2l^2},\\
G^i_{~j}=&\frac{2z(z+n-3)+(n-2)(n-3)}{2l^2}\delta^i_{~j}.
\end{align} 
In general relativity, the corresponding matter field has the diagonal form as $T^\mu_{~~\nu}=\mbox{diag}(-\mu,p_{\rm r},p_{\rm t},p_{\rm t},\cdots)$. 
The physical interpretations of $\mu$, $p_{\rm r}$ and $p_{\rm t}$ are the energy density, radial pressure
and tangential pressure, respectively. 
The weak energy condition
(WEC) implies $\mu \ge 0$, $p_{\rm r}+\mu \ge 0$, and $p_{\rm t}+\mu
\ge 0$, while the dominant energy condition (DEC) implies $\mu \ge
0$, $-\mu \le p_{\rm r} \le \mu$, and $-\mu \le p_{\rm t} \le \mu$.
The null energy condition (NEC) implies $p_{\rm r}+\mu \ge 0$, and
$p_{\rm t}+\mu \ge 0$~\cite{carroll}. Note that DEC implies WEC
and WEC implies NEC.
These quantities for the Lifshitz spacetime (\ref{lifshitz-n}) are given by 
\begin{eqnarray*}
\mu&=&-\frac{(n-1)(n-2)}{16\pi G_nl^2}, \label{mu} \\
p_{\rm r}&=&\frac{(n-3+2z)(n-2)}{16\pi G_nl^2}, \label{pr} \\
p_{\rm t}&=&\frac{2z(z+n-3)+(n-2)(n-3)}{16\pi G_nl^2}. \label{pt}
\end{eqnarray*}
Because of $\mu<0$, DEC and WEC are violated.
However, NEC can be satisfied depending on $z$.
We obtain
\begin{eqnarray*}
\mu+p_{\rm r}&=&\frac{(z-1)(n-2)}{8\pi G_nl^2}, \\
\mu+p_{\rm t}&=&\frac{(z-1)(z+n-2)}{8\pi G_nl^2},
\end{eqnarray*}
and hence NEC is satisfied if and only if $z\ge 1$.
}

The metric (\ref{lifshitz-n}) can be written as
\begin{align}
ds^2=&\frac{r^{2z}}{l^{2z}}(-dt^2+dr_\ast^2)+r^2dx^idx_i,\label{lifshitz-n2}\\
dr_\ast:=&(r/l)^{-1-z}dr.
\end{align} 
If $r=$constant with constant $x^i$ corresponds to finite (infinite) $r_\ast$, it is non-null (null) in the Penrose diagram for the spacetime region covered by this coordinate system.
Therefore, $r=0$ is null for $z\ge 0$ and timelike for $z<0$, while $r=\infty$ is null for $z\le 0$ and timelike for $z>0$.

{Let us clarify the properties of these boundaries, namely $r=0$ and $r=\infty$, depending on the value of $z$.
Consider a radial null geodesic $\gamma$ given by $(t,r)=(t(\lambda),r(\lambda))$ with constant $x^i$, where $\lambda$ 
is an affine parameter.
Its tangent vector is given by $k^\mu=(dt/d\lambda,dr/d\lambda)$.
Since there is a Killing vector $\xi^\mu=\partial/\partial t$ in the Lifshitz spacetime, $C_{(t)}:=k_\mu \xi^\mu$ is conserved along a geodesic.
This equation is written for $\gamma$ as
\begin{align}
C_{(t)}=-\frac{r^{2z}}{l^{2z}}\frac{dt}{d\lambda}.\label{geo1}
\end{align} 
Since $\gamma$ is a radial null geodesic, we have 
\begin{align}
0=-\frac{r^{2z}}{l^{2z}}\biggl(\frac{dt}{d\lambda}\biggl)^2+\frac{l^2}{r^2}\biggl(\frac{dr}{d\lambda}\biggl)^2.\label{geo2}
\end{align} 
Equations~(\ref{geo1}) and (\ref{geo2}) give
\begin{align}
\frac{dr}{d\lambda}=\pm C_{(t)}\frac{r^{-z+1}}{l^{-z+1}},
\end{align} 
which is integrated to obtain
\begin{align}
\pm C_{(t)}(\lambda-\lambda_0)=&\left\{
\begin{array}{ll}
\displaystyle{r^{z}/(zl^{z-1})},\quad \mbox{for} \quad z\ne 0\\
\displaystyle{l\ln|r|},\quad \mbox{for} \quad z= 0
\end{array}
\right.\label{geo-sol}
\end{align}
where $\lambda_0$ is constant.
Null infinity is characterized by $|\lambda|=\infty$.
If $\lambda$ is finite somewhere, it is extendable.
Therefore, for $z>0$, $r=0$ is the extendable symmetric center and $r=\infty$ is infinity along $\gamma$.
In contrast, for $z<0$, $r=0$ is infinity and $r=\infty$ is an extendable boundary.
For the special case with $z=0$, both $r=0$ and $r=\infty$ are infinity.
}

\subsection{Lifshitz solution without potential}
In the case without a potential ($V\equiv 0$), there is the Lifshitz vacuum solution;
\begin{align}
\phi=&\frac{\phi_0}{r^{z+n-2}},\label{scalar-ex2}\\
\omega=&-\frac{2z^2+2(n-2)z+(n-1)(n-2)}{(z+n-2)^2} \label{omega2},
\end{align} 
where $\phi_0$ is a positive constant and $z\ne 2-n$.
{$l$ is arbitrary in this solution.
In contrast, $\phi_0$ can be set to one by rescaling of the coordinates.}

Throughout this paper, we assume $z\ne 2-n$.
$z$ can be written in terms of $\omega$ as
\begin{align}
z=&\frac{-(\omega+1)(n-2)\pm\sqrt{(n-2)(\omega-\omega n-n)}}{\omega+2} \label{z}.
\end{align} 

$\omega \le -n/(n-1)$ is necessary for $z$ to be real, with the equality holding for $z=1$.
Combined this reality condition for $z$ with the condition (\ref{omega-eq}) for the theory to be physically sensible, we find that $\omega$ must satisfy
\begin{align}
-\frac{n-1}{n-2}\le \omega \le -\frac{n}{n-1}.\label{omega3}
\end{align}
This range of $\omega$ corresponds to $0<z \le 2(n-2)/(n-3)$.
In appendix A, we show that this solution is contained within a larger class of vacuum solutions, which is a generalization of the Brans solution in the four-dimensional spherically symmetric spacetime. 
In appendix B, we show that there is no Lifshitz solution in the Einstein frame with a real scalar field.

Here we explain, by showing a black hole no-hair theorem, which are the obstructions for the existence of an asymptotically Lifshitz black hole spacetime is without potential.
We consider the following general static plane-symmetric spacetime 
\begin{align}
ds^2=-\frac{r^{2z}}{l^{2z}}g(r)e^{2\delta(r)}dt^2+\frac{l^2}{r^2}g(r)^{-1}dr^2+r^2dx^idx_i.\label{general}
\end{align} 
In the asymptotically Lifshitz spacetime, $g(r)\to 1$ and $\delta(r)\to 0$ are the boundary conditions to be satisfied for $r\to \infty$. 
The scalar field equation (\ref{beq2}) in vacuum without potential is written as
\begin{align}
\frac{d}{d r}\biggl(\frac{r^{z+1}}{l^{z+1}}e^{\delta(r)}r^{n-2}g(r)\frac{d\phi}{d r}\biggl)=0.
\end{align} 
Multiplying the above equation by $\phi$, we can rewrite it as
\begin{align}
\frac{d}{d r}\biggl(\frac{r^{z+1}}{l^{z+1}}e^{\delta(r)}r^{n-2}g(r)\phi\frac{d\phi}{dr}\biggl)=\frac{r^{z+1}}{l^{z+1}}e^{\delta(r)}r^{n-2}g(r)\biggl(\frac{d\phi}{d r}\biggl)^2.
\end{align} 
Integrating the above equation from the outermost Killing horizon $r=r_{\rm h}$ to $r=\infty$, we obtain
\begin{align}
&\biggl(\frac{r^{z+1}}{l^{z+1}}e^{\delta(r)}r^{n-2}g(r)\phi\frac{d\phi}{d r}\biggl)\biggl|_{r=\infty}-\biggl(\frac{r^{z+1}}{l^{z+1}}e^{\delta(r)}r^{n-2}g(r)\phi\frac{d\phi}{d r}\biggl)\biggl|_{r=r_{\rm h}} \nonumber \\
&~~~~~~~~~~~~~~~~~~~~~=\int_{r_{\rm h}}^{\infty} \biggl\{\frac{r^{z+1}}{l^{z+1}}e^{\delta(r)}r^{n-2}g(r)\biggl(\frac{d\phi}{d r}\biggl)^2\biggl\}dr.
\end{align}
Here we assume (i) $\phi$ and its derivative are finite on the Killing horizon and (ii) $\phi\simeq O(r^{-p})$ with $p>(z+n-2)/2$ for $r\to \infty$.
Since the Killing horizon is defined by $g(r_{\rm h})=0$ with finite $\delta(r_{\rm h})$, the second term on the left-hand side is zero under the assumption (i).
The first term on the left-hand side is zero under the assumption (ii).
Since the right-hand side is positive semi-definite in the domain considered, where $g(r)\ge 0$, we conclude that $\phi$ is constant in this domain of $r$.

Suppose a black hole solution asymptotic to our vacuum Lifshitz solution (\ref{lifshitz-n}) with (\ref{scalar-ex2}) at $r\to \infty$.
Since the power in Eq.~(\ref{scalar-ex2}) satisfies the condition (ii) for $z>2-n$, the only possibility for such a black hole is that $\phi$ or its derivative is diverging at the Killing horizon.
In a scalar-tensor theory, there certainly exists such a black hole solution.
For example, in the Bocharova-Bronnikov-Melnikov-Bekenstein spherically symmetric black hole solution for a conformally coupled scalar field~\cite{Bocharova-Bronnikov-Melnikov,Bekenstein:1975ts}, the scalar field is diverging on the horizon but the geometric quantities are all finite there.
(This solution has been generalized up to the Pleba{\'n}ski-Demia{\'n}ski-type metric until now~\cite{conformalBH}.)

In the next subsection, we explicitly construct an asymptotically Lifshitz black hole solution with analytic $\phi$ on the horizon in the presence of a potential.
({In appendix B, we show that several known solutions represent asymptotically Lifshitz naked singularities.})

\subsection{Asymptotically Lifshitz solution with power-law potential}
Inspired by the Lifshitz vacuum solution obtained in the previous subsection, we derive an exact asymptotically Lifshitz solution in the case with potential ($V\ne 0$).
We consider the following metric
\begin{align}
ds^2=-\frac{r^{2z}}{l^{2z}}f(r)dt^2+\frac{l^2}{r^2}f(r)^{-1}dr^2+r^2dx^idx_i
\label{lifshitz3}
\end{align} 
and assume that the form of the scalar field (\ref{scalar-ex2}) and the relation (\ref{omega2}) between $z$ and $\omega$ remain the same.
Then, Eq.~(\ref{eq4}) is integrated {for $z \ne 1$} to give
\begin{align}
f(r)={\bar k}-\frac{M}{r^{2z-2}},
\end{align}
where ${\bar k}$ and $M$ are constants.
The Ricci scalar in this spacetime is given by
\begin{align}
R=&\frac{(n-1)(n-2z+2)Mr^{2-2z}-{\bar k}[2z(z+n-2)+(n-1)(n-2)]}{l^2},\label{Rscalar}
\end{align}
which is constant for $z=(n+2)/2$.
{For the AdS case with $z=1$, Eq.~(\ref{eq4}) is integrated to give
\begin{align}
f(r)={\bar k}-M\ln |r|
\end{align}
and it does not represent an asymptotically AdS black hole.
The solution for $z=1$ with a Maxwell field is presented in Appendix~\ref{app:z=1}.
Hereafter we assume $z \ne 1$.
}

We are interested in the case with ${\bar k}>0$, in which there can be an outer Killing horizon.
{Here it is shown that we can set ${\bar k}=1$ $\phi_0=1$ simultaneously without loss of generality by rescaling of the coordinates and resorting to the redefinitions of $l$ and $M$.
By the coordinate transformations $t={\bar k}^{(z-1)/2}p^{-z}{\bar t}$, $r=p{\bar r}$, and $x^i=p^{-1}{\bar x}^i$ with $p:=\phi_0^{1/(z+n-2)}$, the Brans-Dicke scalar field (\ref{scalar-ex2}) and the line element become
\begin{align}
\phi=&1/{\bar r}^{z+n-2}, \label{323} \\
ds^2=&-\frac{{\bar k}^{z}{\bar r}^{2z}}{l^{2z}}\biggl(1-\frac{M}{{\bar k}(p{\bar r})^{2z-2}}\biggl)d{\bar t}^2+\frac{l^2}{{\bar k}{\bar r}^2}\biggl(1-\frac{M}{{\bar k}(p{\bar r})^{2z-2}}\biggl)^{-1}d{\bar r}^2+{\bar r}^2d{\bar x}^id{\bar x}_i. \label{324}
\end{align} 
By the reparametrizations ${\bar M}:=M/({\bar k}p^{2z-2})$ and ${\bar l}^2:=l^2/{\bar k}$, the line element finally becomes
\begin{align}
ds^2=-\frac{{\bar r}^{2z}}{{\bar l}^{2z}}\biggl(1-\frac{{\bar M}}{{\bar r}^{2z-2}}\biggl)d{\bar t}^2+\frac{{\bar l}^2}{{\bar r}^2}\biggl(1-\frac{{\bar M}}{{\bar r}^{2z-2}}\biggl)^{-1}d{\bar r}^2+{\bar r}^2d{\bar x}^id{\bar x}_i.
\end{align} 
}

Equations~(\ref{eq1}) (or equivalently Eq.~(\ref{eq2})) and (\ref{eq3}) provide the conditions for the potential
\begin{align}
(n-2)\phi\frac{dV}{d\phi}-nV=&\frac{4z(z-1)(nz-2n+4-3z)\phi_0 M r^{-n-3z+4}}{(z+n-2)l^2},\\
-\frac{V}{2\phi}=&-\frac{z(z-1)Mr^{2-2z}}{l^2}.
\end{align}
{Using Eq.~(\ref{scalar-ex2}), we can show that these equations are indeed satisfied for the following potential
\begin{align}
V(\phi)=V_0\biggl(\frac{\phi}{\phi_0}\biggl)^{(3z+n-4)/(z+n-2)}=:V_{\rm M}, \label{VM}
\end{align}
where constant $V_0$ is given by
\begin{align}
V_0=\frac{2z(z-1)\phi_0M}{l^2}. \label{V0}
\end{align}

This is an exact solution in Brans-Dicke theory with potential (\ref{VM}) in the $n$ spacetime dimensions.
The theory contains two parameters, $\omega$ and $V_0$.
It is realized in the following manner that this is a one-parameter family of solutions.
Since the value of $z$ is determined in terms of $\omega$ and $n$ by Eq.~(\ref{z}) and $\phi_0$ and ${\bar k}$ can be set to one simultaneously as explained, they are not free parameters in the solution.
On the other hand, either $l$ or $M$ can be treated as a free parameter in the solution because Eq.~(\ref{V0}) gives a relation between them.
In the following argument, it is simpler to take $M$ as an independent parameter.
A very curious property of our solution is that we cannot take the limit $M \to 0$ because it implies $l \to 0$ bv the condition (\ref{V0}), in other words, there is no limit to the exact Lifshitz spacetime.
In this sense, we could say that our black hole does not have any background spacetime. 
}

This solution represents an asymptotically locally Lifshitz black hole for $z>1$ and $M>0$.
Then the potential is positive definite and there is a single outer Killing horizon at $r=M^{1/(2z-2)}=:r_{\rm h}$, which is an event horizon.
The Ricci scalar (\ref{Rscalar}) diverges at $r=0$ for $z>1$ and $M \ne 0$.
Therefore, there is a curvature singularity at the origin $r=0$.
The spacetime can be written as
\begin{align}
ds^2=&\frac{r^{2z}}{l^{2z}}f(r)(-dt^2+dr_\ast^2)+r^2dx^idx_i,\\
dr_\ast:=&\frac{l^{z+1}}{r^{z+1}}f(r)^{-1}dr.
\end{align} 
Therefore, the curvature singularity at $r=0$ is null for $1<z \le 2$ and non-null for $z>2$.
Because of condition (\ref{omega3}), the asymptotically Lifshitz black hole is realized in the physically sensible theory for $1<z \le 2(n-2)/(n-3)$.
It is noted that our solution with $z=n$ reduces to the solution obtained by Dehghani, Pakravan, and Hendi in~\cite{dph2006} with zero rotation.

\subsection{Thermodynamical properties of the Lifshitz black hole}
In this subsection, we discuss the thermodynamical properties of our black hole with $z>1$.
Let us write down again the metric with ${\bar k}=1$ and the scalar field here;
\begin{align}
ds^2=&-\frac{r^{2z}}{l^{2z}}f(r)dt^2+\frac{l^2}{r^2}f(r)^{-1}dr^2+r^2dx^idx_i,\label{diagonal}\\
f(r)=&1-\frac{M}{r^{2z-2}},\qquad \phi=\frac{\phi_0}{r^{z+n-2}}.
\end{align} 
A Killing horizon of this spacetime is defined by $f(r_{\rm h})=0$, which is solved to give
\begin{align}
M=&r_{\rm h}^{2z-2}.
\end{align} 
{Using Eq.~(\ref{V0}), $l$ is written in terms of $r_{\rm h}$ as
\begin{align}
l= \biggl(\frac{2z(z-1)\phi_0}{V_0}\biggl)^{1/2}r_{\rm h}^{z-1}.
\end{align}
} First, we derive the surface gravity and the temperature of the black hole by the Euclidean method.
With the Euclidean time $\tau:=it$, the metric (\ref{diagonal}) becomes
\begin{align}
ds^2=&\frac{r^{2z}}{l^{2z}}f(r)d\tau ^2+\frac{l^2}{r^2}f(r)^{-1}dr^2+r^2dx^idx_i.
\end{align} 
Near the horizon $r=r_{\rm h}$, we obtain 
\begin{align}
f(r)\simeq &(2z-2)r_{\rm h}^{-1}(r-r_{\rm h})
\end{align} 
and therefore the near-horizon metric becomes
\begin{align}
ds^2 \simeq \frac{r_{\rm h}^{2z-1}}{l^{2z}}(2z-2)(r-r_{\rm h})d\tau ^2+\frac{l^2}{(2z-2)r_{\rm h}(r-r_{\rm h})}dr^2+r_{\rm h}^2dx^idx_i.
\end{align} 
By the coordinate transformation 
\begin{align}
r-r_{\rm h}=\frac{(2z-2)r_{\rm h}}{4l^2}x^2,
\end{align} 
we can write the two-dimensional part of the near-horizon metric in the Rindler form as
\begin{align}
ds^2 \simeq & \ dx^2+x^2d(\kappa\tau)^2+r_{\rm h}^2dx^idx_i,\\
\kappa:=&\frac{|z-1|r_{\rm h}^{z}}{l^{z+1}}.
\end{align} 
If we impose periodicity of $\tau$ as $\tau\sim \tau+2\pi/\kappa$, there is no conical singularity at $x=0$, namely on the horizon.
We identify this $\kappa$ as the surface gravity on the horizon.
The temperature of a black hole is defined by $T:=\kappa/(2\pi)$. Namely
\begin{align}
T =& \frac{|z-1|r_{\rm h}^{z}}{2\pi l^{z+1}}, \nonumber \\
=&  \frac{|z-1|}{2\pi} \biggl(\frac{2z(z-1)\phi_0}{V_0}\biggl)^{-(1+z)/2}r_{\rm h}^{1+z-z^2}. \label{342}
\end{align} 
{It is seen that, for the special case with $z=(1+\sqrt{5})/2$, the temperature of the black hole is independent of its size.}

Here, we come to a point that deserves special attention. That is, thermodynamics.
The Lifshitz black hole solution (\ref{323})-(\ref{324}) exhibits a peculiar feature, which is the fact that, as already pointed out,
its parameter $M$ gets fixed in terms of $l$, which ultimately determines the curvature of the asymptotic geometry.
This is associated to the fact that there is no $M \to 0$ limit yielding the Lifshitz geometry.
Consequently, there is no clear notion of a reference background in this setup, and this makes the thermodynamics analysis unclear.
In contrast to what happens with temperature (\ref{342}), which is a well defined geometrical quantity,
the notions of entropy and energy associated to these spacetimes become fuzzy.
Nevertheless, having said that, and taking this important cautionary note into account, one can insist in performing a thermodynamics analysis using the formulas for entropy that hold for Brans-Dicke theory in different scenarios, and then assume the first law of black hole thermodynamics to hold in order to propose a notion of mass.
We assume that the $(n-2)$-dimensional flat submanifold $(K^{n-2},\gamma _{ij})$ is compactified with volume $V_{(n-2)}$.
The Wald entropy $S$~\cite{waldentropy} (being computed on the Killing horizon) in Brans-Dicke theory is given by 
\begin{align}
S=\frac{\phi_{\rm h}}{4}A_{\rm h}=\frac{\phi_0V_{(n-2)}}{4r_{\rm h}^z},
\end{align} 
where $\phi_{\rm h}$ is the value of $\phi$ on the horizon and $A_{\rm h}:=V_{(n-2)}r_{\rm h}^{n-2}$ is the horizon area.
We obtain
\begin{align}
\frac{dS}{dr_{\rm h}}=-\frac{\phi_0zV_{(n-2)}}{4r_{\rm h}^{z+1}}
\end{align} 
and thus a larger (smaller) black hole has smaller (larger) entropy.

It is interesting to compare the thermodynamics of our solution for $n=4$ with that of the four-dimensional analytic 
Lifshitz black hole solution found in reference \cite{BalasubramanianMcGreevy2}. For $n=4$ and $z=2$ the metric of our 
solution coincides with the metric found there. Consequently, the temperature of both black holes, being $T$ a purely 
geometric quantity, agrees. In contrast, the entropy of the black hole solution of \cite{BalasubramanianMcGreevy2} 
substantially differs from the one we found here: While in our case the entropy $S$ is proportional to $r_{\rm 
h}^{-2}$, the entropy found in \cite{BalasubramanianMcGreevy2} turns out to be proportional to $r_{\rm h}^2$ (notice 
that the coordinate $r$ in \cite{BalasubramanianMcGreevy2} is the inverse to the one used here.) This difference, which 
has implications in the thermal (in)stability of the solution, is due to the fact that the two theories are totally 
different and, in particular, because of the non-minimal coupling of the Brans-Dicke field. 

Then, from the results above we obtain
\begin{align}
TdS=-\frac{\phi_0z|z-1|V_{(n-2)}}{8\pi} \biggl(\frac{2z(z-1)\phi_0}{V_0}\biggl)^{-(1+z)/2}r_{\rm h}^{-z^2} dr_{\rm h}.
\end{align} 
Although the existence of the first law happens to be a non-trivial assumption in the Lifshitz spacetime, here we asume it for simplicity in order to propose a notion of {\it mass} for these asymptotically Lifshitz spaces. That is, assuming $dE=TdS$, the energy $E$ can be obtained by simple integration as
\begin{align}
E=&\frac{\phi_0z|z-1|V_{(n-2)}}{8\pi(z+1)(z-1)} \biggl(\frac{2z(z-1)\phi_0}{V_0}\biggl)^{-(1+z)/2}r_{\rm h}^{-(z+1)(z-1)}+E_0, \nonumber \\
=&\frac{\phi_0z|z-1|V_{(n-2)}}{8\pi(z+1)(z-1)} \biggl(\frac{2z(z-1)\phi_0}{V_0}\biggl)^{-(1+z)/2}M^{-(z+1)/2}+E_0,
\end{align} 
where $E_0$ is a constant.
{It is seen that a larger (smaller) black hole has smaller (larger) value of $E$.
Using the quantities above, we can obtain the Smarr-like relation as
\begin{align}
E-E_0=\frac{z}{(z+1)(z-1)}TS.
\end{align} 
The constant $E_0$ corresponds to the value of $E$ for an infinitely large black hole.
For the standard AdS black hole, such a value is fixed in such a way that the background spacetime has vanishing energy.
However in our case, because of the absence of the limit to the exact Lifshitz spacetime, there is no simple criterion to fix $E_0$.
}

Lastly, we derive the thermodynamical quantities.
The heat capacity $C$ is given by
\begin{align}
C:=&\frac{dE}{dT}=\frac{dE}{dr_{\rm h}} \biggl( \frac{dT}{dr_{\rm h}} \biggl)^{-1} \nonumber \\
=&\frac{\phi_0zV_{(n-2)}}{4(z^2-z-1)r_{\rm h}^{z}}.
\end{align}
Positive (negative) heat capacity implies that the black hole is locally thermodynamically stable (unstable). 
{For $z>(<)(1+\sqrt{5})/2$, the heat capacity of our black hole is positive (negative).}
To analyze the global thermodynamical stability one must obtain the free energy of the black hole defined by $F:=E-TS$.
We obtain
\begin{align}
F=&-\frac{\phi_0(z^2-z-1)|z-1|V_{(n-2)}}{8\pi(z+1)(z-1)} \biggl(\frac{2z(z-1)\phi_0}{V_0}\biggl)^{-(1+z)/2}r_{\rm h}^{-(z+1)(z-1)}+E_0.
\end{align} 
{Negative (positive) free energy implies that the black hole is globally thermodynamically stable (unstable). 
However, as expected, the absence of the natural way to fix $E_0$ prevents us to discuss the global thermodynamical stability.
}

\section{Lifshitz black holes with Maxwell field}
In this section, we present a generalized solution in the presence of a Maxwell field.
We obtain the solution with the metric (\ref{lifshitz3}) and the scalar field (\ref{scalar-ex2}) under the relation (\ref{omega2}) between $z$ and $\omega$.

\subsection{Solution for the Maxwell equation}
First, let us solve the Maxwell equation (\ref{max-d}).
The electric Maxwell equation $D_bF^{ab}=0$ can be integrated to give
\begin{align}
A_{a}dy^a=-\frac{Q_{\rm e}}{r^{n-2-z}}dt,\label{A-e}
\end{align}
where $Q_{\rm e}$ is a real constant, from which we obtain (for $n \neq z+2$)
\begin{align}
F_{tr}=-\frac{(n-2-z)Q_{\rm e}}{r^{n-1-z}}, \qquad F_{ab}F^{ab}=-\frac{2(n-2-z)^2l^{2z-2}Q_{\rm e}^2}{r^{2(n-2)}}.
\end{align}
This solution exists in arbitrary dimensions.

On the contrary, the solution for the magnetic Maxwell equation ${\bar D}_jF^{ij}=0$ exists only in even dimensions~\cite{opz2008,mhm2010}.
Writing the metric on the $(n-2)$-dimensional flat space as
\begin{align}
&\gamma_{ij}dz^idz^j=\sum_{\sigma=1}^{(n-2)/2}(d\theta_\sigma^2+d\phi_\sigma^2),
\end{align}
we can show that the following Faraday tensor satisfies the magnetic Maxwell equation ${\bar D}_jF^{ij}=0$~\cite{mhm2010}
\begin{align}
F_{ij}d z^i \wedge d z^j =Q_{\rm m}\sum_{\sigma=1}^{(n-2)/2}(d \theta_\sigma \wedge d \phi_\sigma),
\end{align}
which is consistent with Eq.~(\ref{mag-F}).
The magnetic component of the gauge field is given, for example, by
\begin{align}
A_idz^i=Q_{\rm m}\sum_{\sigma=1}^{(n-2)/2}\theta_\sigma d \phi_\sigma.\label{A-m}
\end{align}
In summary, the electric solution (\ref{A-e}) exists in arbitrary dimensions, while the magnetic solution (\ref{A-m}) exists in even dimensions.
In other words, $Q_{\rm m}=0$ is required in odd dimensions.

\subsection{Asymptotically Lifshitz solutions with Maxwell field}
Now we are ready to obtain the solution with a Maxwell field.
Using Eq.~(\ref{eq3}), we obtain the metric function as
\begin{align}
f(r)=&{\bar k}-\frac{M}{r^{2z-2}}+f_{\rm Q_{\rm e}}(r)+f_{\rm Q_{\rm m}}(r),\\
f_{\rm Q_{\rm e}}(r):=&\left\{
\begin{array}{ll}
\displaystyle{\frac{64\pi l^{2z}(n-2-z)Q_{\rm e}^2}{\phi_0\zeta^2(n-3z)r^{n-2-z}}},\quad \mbox{for} \quad z\ne n/3\\
\displaystyle{-\frac{128\pi(n-3)Q_{\rm e}^2 l^{2n/3}\ln|r|}{3\zeta^2\phi_0r^{2(n-3)/3}}},\quad \mbox{for} \quad z= n/3
\end{array}
\right.\label{fq} \\
f_{\rm Q_{\rm m}}(r):=&\left\{
\begin{array}{ll}
\displaystyle{-\frac{64\pi l^2Q_{\rm m}^2}{\phi_0\zeta^2(3z-8+n)(6-n-z)r^{6-n-z}}},\quad \mbox{for} \quad z\ne (8-n)/3, 6-n \\
\displaystyle{\frac{96\pi l^2Q_{\rm m}^2\ln|r|}{\phi_0\zeta^2(n-5)r^{2(5-n)/3}}},\quad \mbox{for} \quad z=(8-n)/3\\
\displaystyle{-\frac{32\pi l^2Q_{\rm m}^2}{\phi_0\zeta^2(n-5)}\ln|r|}, \quad \mbox{for} \quad z=6-n.
\end{array}
\right.\label{fc}
\end{align}
In a similar manner to the vacuum case, we identify the potential as
\begin{align}
V(\phi)=&V_{\rm M}+V_{\rm Q_{\rm e}}+V_{\rm Q_{\rm m}},\\
V_{\rm Q_{\rm e}}:=&\left\{
\begin{array}{ll}
\displaystyle{\frac{32\pi l^{2z-2}(n-z)(n-z-2)^2Q_{\rm e}^2}{(3z-n)\zeta^2}\biggl(\frac{\phi}{\phi_0}\biggl)^{2(n-2)/(z+n-2)}},\quad \mbox{for} \quad z\ne n/3\\
\displaystyle{-\frac{128\pi l^{2(n-3)/3}(n-3) Q_{\rm e}^2[n(n-3)\ln|\phi/\phi_0|+(2n-3)^2]}{9\zeta^2(2n-3)}\biggl(\frac{\phi}{\phi_0}\biggl)^{3(n-2)/(2n-3)}}, \quad \mbox{for} \quad z= n/3
\end{array}
\right.\label{Vq} \\
V_{\rm Q_{\rm m}}:=&\left\{
\begin{array}{ll}
\displaystyle{-\frac{16\pi [(3n-10)z+(n-2)(n-8)]Q_{\rm m}^2}{(3z-8+n)\zeta^2}\biggl(\frac{\phi}{\phi_0}\biggl)^{4/(z+n-2)}},\quad \mbox{for} \quad z\ne (8-n)/3 \\
\displaystyle{\frac{16\pi Q_{\rm m}^2[2(n-5)(n-8)\ln|\phi/\phi_0|+(n+1)^2(n-6)]}{(n+1)(n-5)\zeta^2}\biggl(\frac{\phi}{\phi_0}\biggl)^{6/(n+1)}}, \quad \mbox{for} \quad z=(8-n)/3.
\end{array}
\right.\label{Vc}
\end{align}
It is observed that $V_{\rm Q_{\rm e}}=0$ is satisfied for $z=n$.
Namely, the electrically charged solution is obtained with a simple quadratic potential.
(Our solution with $z=n$ and $Q_{\rm m}=0$ reduces to the solution obtained in~\cite{dph2006} with zero rotation.)
{Also, $V_{\rm Q_{\rm m}}=0$ is realized for satisfied for $z=(n-2)(8-n)/(3n-10)$.
The condition (\ref{omega3}) requires $0<z \le 2(n-2)/(n-3)$, which is compatible with $V_{\rm Q_{\rm e}}=0$ or/and $V_{\rm Q_{\rm m}}=0$ for $n=4$ and the corresponding values of $z$ and $\omega$ are $z=4$ and $\omega=-3/2$ in both cases.
(We should remember that the electric field is pure gauge for $n=3$ and $Q_{\rm m}\equiv 0$ for odd $n$.)
In this case, namely for $n=4$ and $\omega=-3/2$, the parameters $Q_{\rm e}$ and $Q_{\rm m}$ are independent parameters, however, the spacetime is not asymptotically Lifshitz for $r\to \infty$.

Now let us consider the case with non-zero $V_{\rm Q_{\rm e}}$ and $V_{\rm Q_{\rm m}}$.
In comparison with the neutral case, there apparently appeared two additional parameters in the solution, $Q_{\rm e}$ and $Q_{\rm m}$.
However, this is still a one-parameter family of solutions shown as follows.
It is realized that we can set both $\phi_0(>0)$ and ${\bar k}(>0)$ to one simultaneously by rescaling the coordinates and redefinition of $l^2$, $M$, $Q_{\rm e}^2$, and $Q_{\rm m}^2$ as explicitly performed in the previous section in the neutral case.
($Q_{\rm e}^2$ and $Q_{\rm m}^2$ should be redefined as $Q_{\rm e}^2={\bar Q}_{\rm e}^2p^{2(n-2)}{\bar k}^{1-z}$ and $Q_{\rm m}^2={\bar Q}_{\rm m}^2p^4$.)
As also explained, Eq.~(\ref{VM}) gives a relation between the constant in $V_{\rm M}(\phi)$ (namely $V_0$) and the constants in the solution, $l$ and $M$.
In contrast, Eq.~(\ref{Vc}) gives a relation between two constants in the theory, namely the constant in $V_{\rm Q_{\rm m}}(\phi)$ (say $V_2$) and $\zeta$, and the constant in the solution, $Q_{\rm m}$.
Therefore, the value of $Q_{\rm m}$ is totally fixed by the parameters in the theory; $\omega$, $\zeta$, and $V_2$.
On the other hand, Eq.~(\ref{Vq}) gives a relation between two constants in the theory, the constant in $V_{\rm Q_{\rm e}}(\phi)$ (say $V_1$) and $\zeta$, and two constants in the solution, $l$ and $Q_{\rm e}$.
In summary, while the value of $Q_{\rm m}$ is fixed by the parameters in the theory, there are two relations between $l$, $M$, and $Q_{\rm e}$.
One can take one of either $l$, $M$, or $Q_{\rm e}$ as a free parameter in the solution.
If we take $M$ as a free parameter, the value of $l$ is determined by Eq~(\ref{V0}) and subsequently the value of $Q_{\rm e}$ is determined by Eq.~(\ref{Vq}).
As in the neutral case, there is no limit to the exact Lifshitz solution.
}

As we explained, $Q_{\rm m}=0$ is required in odd dimensions.
Since the potential form is quite complicated, let us clarify whether the solution with $M=Q_{\rm m}=0$ in arbitrary dimensions or $M=Q_{\rm e}=0$ in even dimensions can be an asymptotically Lifshitz black hole in the physical theory or not.
We are interested in the case with positive ${\bar k}$ and then we set ${\bar k}=1$ without loss of generality as in the vacuum case.
First, it is noted that the condition for the physical theory (\ref{omega-eq}) requires the positivity of $z$.

First, let us consider the case with $M=Q_{\rm m}=0$ in arbitrary dimensions.
Here we don't consider the case with $n=3$ and $z=1$ since it gives nothing but the AdS solution with a pure gage of $A_\mu$.
The spacetime is asymptotically Lifshitz for $z<n-2$.
For $z \ne n/3$, the potential is non-negative for $n/3<z \le n$ with equality giving $V_{\rm Q_{\rm e}}=0$.
For $z=n/3$, on the other hand, the potential becomes negative for a sufficiently small value of $r$.
Therefore, the fall-off condition and non-negative potential for any $\phi$ are achieved only for $n\ge 4$ and 
$n/3<z<n-2$.
Then, the coefficient of $Q_{\rm e}^2$ in $f_{\rm Q_{\rm e}}(r)$ is negative and the spacetime represents an asymptotically Lifshitz black hole.
However, using the condition (\ref{omega3}), we can show that this black hole is realized in the physical theory only for $4 \le n \le 7$, in which $z$ must satisfy $n/3<z \le 2(n-2)/(n-3)$.

Next we consider the case with $M=Q_{\rm e}=0$ in even dimensions.
The physical theory condition (\ref{omega3}) requires $z>0$.
Then, the spacetime is asymptotically Lifshitz only for $n=4$ with $0<z<2$.
In this range of $z$, the potential is positive for $4/3 < z <2$.
For $z=4/3$, on the other hand, the potential becomes negative for a sufficiently large value of $r$.
Then, the fall-off condition and non-negative potential for any $\phi$ are achieved only for $4/3 < z <2$.
Then, the coefficient of $Q_{\rm m}^2$ in $f_{\rm Q_{\rm m}}(r)$ is negative and the spacetime represents an asymptotically Lifshitz black hole.
In this range of $z$, the theory is physical.
Our result is summarized in Table~\ref{table:th}.

\begin{table}[h]
\begin{center}
\caption{\label{table:th} Conditions for the asymptotically Lifshitz black hole with either one of $M$, $Q_{\rm e}$, or $Q_{\rm m}$ is non-zero in the physical theory with positive potential.
}
\begin{tabular}{l@{\qquad}c@{\qquad}c}
\hline \hline
 Parameters & Conditions    \\\hline
$M\ne 0$ with $Q_{\rm e}=Q_{\rm m}=0$ & $M>0$ and $1<z \le 2(n-2)/(n-3)$   \\ \hline
$Q_{\rm e}\ne 0$ with $M=Q_{\rm m}=0$ & $4 \le n \le 7$ and $n/3<z \le 2(n-2)/(n-3)$  \\ \hline
$Q_{\rm m}\ne 0$ with $M=Q_{\rm e}=0$ & $n=4$ and $4/3 < z <2$   \\ 
\hline \hline
\end{tabular}
\end{center}
\end{table} 

\section{Summary}

In this paper, we have constructed an exact asymptotically Lifshitz black hole solution in Brans-Dicke theory both with and without a Maxwell field in arbitrary dimensions.
The dynamical exponent $z$ is given in terms of the Brans-Dicke parameter $\omega$ and the number of spacetime dimensions $n$.
In the vacuum case, a power-law potential for the Brans-Dicke scalar field is needed.
For the solution to represent an asymptotically Lifshitz black hole in the physical theory, $1<z \le 2(n-2)/(n-3)$ is required, which corresponds to the negative range of values of $\omega$.
We have computed the Hawking temperature of the Lifshitz black hole solution, and we have discussed the problems encountered in analyzing its thermodynamics. Although a notion of mass for these asymptotically Lifshitz spaces can be proposed by assuming the generalized first law, a full understanding of their thermodynamics is still missing.
We have successfully generalized the solution in presence of a Maxwell field.

There are several open problems on the general relativity side.
We have not clarified all the possible black hole configurations depending on the three parameters.
For instance, it would be interesting to check whether there exists an extremal black hole configuration.
Other open problem is about the calculation of the mass: In our study of the thermodynamical properties of a black 
hole, we have assumed the first law to hold.
However, this is a highly non-trivial assumption in the asymptotically Lifshitz spacetime.
To check whether it is satisfied by defining a global mass as a conserved charge in the asymptotically Lifshitz 
spacetime is an important task to be addressed.
Lastly, although it is still not clear, the dynamical stability of the black hole is an important task to be addressed. 
Regarding this issue of stability, it is worth noticing that the condition 
(\ref{omega3}) we used to impose restrictions on $z$ actually corresponds to the no-ghost condition about flat space. 
In fact, the property of 
being free of ghosts is a 
background dependent one, and the fact of having considered (\ref{omega3}) has to be regarded merely as a criterion to 
select a physically sensible model. So, the question still remains as to what are the restrictions on the parameters 
and 
which are the asymptotic falling-off conditions to be imposed for the theory to be free of ghost instabilities about 
Lifshitz spaces.

On the other hand, in a string theory context, in which the holographic realizations appear naturally, a question that 
still remains is that about the embedding of the Lifshitz 
black hole solution in a stringy framework. As emphasized in \cite{BalasubramanianMcGreevy2}, a problem with this type 
of setup is that Lifshitz black holes usually appear in models with {\it ad hoc} matter content which are hardly 
natural from the string theory point of view. Having shown that analytic Lifshitz black holes do appear in a particular 
kind of scalar-tensor scenario is certainly interesting and may provide some insight to look at the problem. 
Nevertheless, something like an analytic Lifshitz black hole in a string model is still missing.

\subsection*{Acknowledgements}
The authors thank C.~Mart\'{\i}nez and M.~Nozawa for useful comments. 
H.M. thanks M.~Hassa\"{\i}ne and K.~Copsey for stimulating conversations. G.G. thanks E.~Ay\'on-Beato, A.~Garbarz, and 
M.~Hassa\"{\i}ne for 
previous collaboration in related topics. This work has been partially funded by the Fondecyt grants 1100328, 1100755 
(HM), and by the Conicyt grant "Southern Theoretical Physics Laboratory" ACT-91, and by grants APNCyT, CONICET and 
UBACyT. 
This work was also partly supported by the JSPS Grant-in-Aid for Scientific Research (A) (22244030).
The Centro de Estudios Cient\'{\i}ficos (CECs) is funded by the Chilean Government through the Centers of Excellence 
Base Financing Program of Conicyt.

\appendix

\section{Generalized Brans solution contains (asymptotically) Lifshitz solutions}
In this appendix, we show that our Lifshitz vacuum solution is contained by a larger class of vacuum solutions without potential.
In~\cite{brans1962}, Brans presented four forms (I--IV) of an exact spherically symmetric static vacuum solution in Brans-Dicke theory without potential in four dimensions.
There is a direct coordinate transformation between the forms I and II.
There is also a direct coordinate transformation between the forms III and IV, which correspond to the case with $\lambda=0$ or $\Lambda=0$ in the forms I and II, respectively.
(See also~\cite{bs2005}.)
Here we show that the Brans solution and its generalization for arbitrary $k$ and $n$ contain exact or asymptotically Lifshitz solutions but they do not represent a black hole.
{The generalized Lifshitz spacetime for arbitrary $k$ is defined by 
\begin{align}
ds^2=-\frac{r^{2z}}{l^{2z}}dt^2+\frac{l^2}{r^2}dr^2+r^2\gamma_{ij}dz^idz^j,\label{lifshitz-k}
\end{align} 
where $\gamma_{ij}dz^idz^j$ is a unit metric on the maximally symmetric space with sectional curvature $k=0,\pm 1$.
}

\subsection{Generalized Brans solution without potential}
The Brans solution with $\Lambda\ne 0\ne \lambda$ can be generalized for arbitrary $n$ and $k$ with a different parametrization and a coordinate system as
\begin{align}
ds^2=&-h(\rho)^{a+1}dt^2+h(\rho)^{b-1}d\rho^2+\rho^2h(\rho)^b\gamma_{ij}dz^idz^j,\label{brans}\\
\phi=&\phi_0h(\rho)^{-[a+(n-3)b]/2},\\
h(\rho)=&k-\frac{m}{\rho^{n-3}},\\
\omega=&-\frac{2a^2+2[(n-3)b+1]a+(n-2)(n-3)b^2-2b}{[a+(n-3)b]^2},\label{omega-brans}
\end{align}
where $a$, $b$, and $m$ are constants.
The expression of $\omega$ (\ref{omega-brans}) is singular at $a=-(n-3)b$.
If we set $a=-(n-3)b$ in the metric, with which $\phi$ is constant, the field equations give $(a,b)=(-2,2/(n-3))$ or $(0,0)$ with arbitrary $\omega$ for $n \ne 3$.
Indeed, this solution is equivalent to the Lifshitz solution in the case with $k=0$ and $b=(n-1)/(n-3)$ for $n\ge 4$.
The dynamical exponent $z$ is related to the parameter $a$ as $z= a+1$.
{This is confirmed directly by the coordinate transformation $r=\rho^{(3-n)/2}$.}

{However, the Lifshitz solution in three dimensions is not contained by this class of solutions.
In three dimensions ($n=3$), the solution for the metric (\ref{brans}) with non-constant $h(\rho)$ is given by
\begin{align}
\phi=&\phi_0h(\rho)^{-(2+a)/2},\\
h(\rho)=&h_0\rho^{-4/[\omega(a+2)^2+2b+2(a+1)(a+2)]},
\end{align}
where $a$ and $b$ are arbitrary and $\omega$ must satisfy $\omega \ne -2\{b+(a+1)(a+2)\}/(a+2)^2$.
$h_0$ is an integration constant.
The following parameter set gives the three-dimensional Lifshitz spacetime;
\begin{align}
a=z-1,\quad b=0,\quad \omega =-\frac{2(z^2+z+1)}{(1+z)^2}.
\end{align}

Actually, the generalized Brans spacetime (\ref{brans}) contains an asymptotically Lifshitz spacetime as an special case.
Defining $r:=(-m)^{b/2}\rho^{[2-b(n-3)]/2}$ as well as $l$ such that 
\begin{align}
l^{2(n-3)/[b(n-3)-2]}=\frac{4}{[2-b(n-3)]^2}(-m)^{2/[b(n-3)-2]},
\end{align}
we rewrite the generalized Brans solution (\ref{brans}) in the form of
\begin{align}
ds^2=&-\frac{r^{2(a+1)(n-3)/[b(n-3)-2]}}{l^{2(a+1)(n-3)/[b(n-3)-2]}}A(r)^{a+1}d{\tilde t}^2+\frac{l^{2(n-3)/[b(n-3)-2]}}{r^{2(n-3)/[b(n-3)-2]}}A(r)^{b-1}dr^2+r^2A(r)^b\gamma_{ij}dz^idz^j,\\
\phi=&{\tilde\phi}_0\frac{l^{(n-3)[a+(n-3)b]/[b(n-3)-2]}}{r^{(n-3)[a+(n-3)b]/[b(n-3)-2]}}A(r)^{-[a+(n-3)b]/2},\\
A(r):=&1+\frac{k[2-b(n-3)]^2l^{2(n-3)/[b(n-3)-2]}}{4r^{2(n-3)/[b(n-3)-2]}},
\end{align}
where we defined
\begin{align}
{\tilde t}:=&\biggl(\frac{[2-b(n-3)]^2}{4}\biggl)^{-(a+1)/2}t,\\
{\tilde\phi}_0:=&\phi_0\biggl(\frac{[2-b(n-3)]^2}{4}\biggl)^{[a+(n-3)b]/2}.
\end{align}
Therefore, this spacetime is asymptotically Lifshitz for $r\to \infty$ if
\begin{align}
b=\frac{n-1}{n-3} \label{b-brans}
\end{align}
and then the dynamical exponent is given by
\begin{align}
z=a+1.
\end{align}
This solution can be written in a more friendly form as
\begin{align}
ds^2=&-\frac{r^{2z}}{l^{2z}}A(r)^{z}d{\tilde t}^2+\frac{l^{2}}{r^{2}}A(r)^{b-1}dr^2+r^2A(r)^b\gamma_{ij}dz^idz^j,\\
\phi=&{\tilde\phi}_0\frac{l^{z+n-2}}{r^{z+n-2}}A(r)^{-(z+n-2)/2},\\
A(r):=&1+\frac{k(n-3)^2l^{2}}{4r^{2}},
\end{align}
where $n\ne 3$. 
It is seen that there is a zero in $A(r)$ for $k=-1$.
However, it is shown to be a curvature singularity.
The Ricci scalar of the generalized Brans spacetime (\ref{brans}) is given by
\begin{align}
R=-\frac{(n-3)^2m^2[2a^2+2\{(n-3)b+1\}a+(n-2)(n-3)b^2-2b]}{4\rho^{2n-4}h(\rho)^{1+b}}.
\end{align}
Since we have $b=(n-1)/(n-3)>0$, {the only possibility to avoid the singularity at $h(\rho)=0$ is $n^2-n+2na+2a^2=0$.
However, it is not satisfied for real $a$.
Therefore, it is concluded that this class of generalized Brans solution represents not an asymptotically Lifshitz black hole but an asymptotically Lifshitz naked singularity.
}

\subsection{Generalized Brans solution with potential}

Another solution we obtain in the presence of potential is
\begin{align}
ds^2=&-H(\rho)^{a+1}dt^2+H(\rho)^{a-1}d\rho^2+\rho^2H(\rho)^a\gamma_{ij}dz^idz^j,\label{brans-pot}\\
\phi=&\phi_0H(\rho)^{-(n-2)a/2},\\
H(\rho)=&k-\frac{m}{\rho^{n-3}}-\frac{2\Lambda }{(n-1)(n-2)}\rho^2,\\
\omega=&-\frac{n-1}{n-2},\label{omega-brans2}\\
V(\phi)=&2\phi_0\Lambda\biggl(\frac{\phi}{\phi_0}\biggl)^{n/(n-2)}
\end{align}
where $a$, $m$, and $\Lambda$ are constants.
The metric becomes the Schwarzschild-(A)dS metric for $a=0$ with a constant scalar field.
We don't consider the case with $a=0$ and we assume $\Lambda\ne 0$.
It is seen that there can be zeros in $H(r)$ if $km\ne 0$.
The Ricci scalar of the spacetime (\ref{brans-pot}) blows up at $H(\rho)=0$ for $km\ne 0$ with $a>-1~(a\ne 0)$.

The spacetime (\ref{brans-pot}) is not asymptotically Lifshitz for $\rho \to \infty$ if $a \ne 0$.
On the other hand, it can be asymptotically Lifshitz for $\rho \to 0$.
With the same transformation as in the previous subsection, we write the solution as
\begin{align}
ds^2=&-\frac{r^{2(a+1)(n-3)/[a(n-3)-2]}}{l^{2(a+1)(n-3)/[a(n-3)-2]}}B(r)^{a+1}d{\tilde t}^2+\frac{l^{2(n-3)/[a(n-3)-2]}}{r^{2(n-3)/[a(n-3)-2]}}B(r)^{a-1}dr^2+r^2B(r)^a\gamma_{ij}dz^idz^j,\\
\phi=&{\tilde\phi}_0\frac{l^{(n-3)(n-2)a/[a(n-3)-2]}}{r^{(n-3)(n-2)a/[a(n-3)-2]}}B(r)^{-(n-2)a/2},\\
B(r):=&1+\frac{k[2-a(n-3)]^2l^{2(n-3)/[a(n-3)-2]}}{4r^{2(n-3)/[a(n-3)-2]}}-\biggl(\frac{[2-a(n-3)]^2}{4}\biggl)^a\frac{{\tilde\Lambda}l^{2a(n-3)/[a(n-3)-2]}}{r^{4/[a(n-3)-2]}},
\end{align}
where
\begin{align}
{\tilde\Lambda}:=\frac{2\Lambda}{(n-1)(n-2)}.
\end{align}
This spacetime is asymptotically Lifshitz spacetime for $r\to \infty$ if 
\begin{align}
a=\frac{n-1}{n-3}
\end{align}
with the dynamical exponent $z$ such that 
\begin{align}
z=a+1=\frac{2(n-2)}{n-3}.
\end{align}
The solution in a friendly form is
\begin{align}
ds^2=&-\frac{r^{2z}}{l^{2z}}B(r)^{z}d{\tilde t}^2+\frac{l^{2}}{r^{2}}B(r)^{a-1}dr^2+r^2B(r)^a\gamma_{ij}dz^idz^j,\\
\phi=&{\tilde\phi}_0\frac{l^{(n-2)a}}{r^{(n-2)a}}B(r)^{-(n-2)a/2},\\
B(r):=&1+\frac{k(n-3)^2l^{2}}{4r^{2}}-\biggl(\frac{(n-3)^2}{4}\biggl)^a\frac{{\tilde\Lambda}l^{2a}}{r^{4/(n-3)}}.
\end{align}
Since now we have $a=(n-1)/(n-3)>0$, $B(r)=0$ is not a horizon but a curvature singularity.
As a result, this solution represents an asymptotically Lifshitz naked singularity, too.
}

\section{No Lifshitz solution in the Einstein frame with a real scalar field}
In this appendix, we show the incompatibility of the Lifshitz solution in the Einstein frame with a real scalar field.
Let us consider the following action in the Einstein frame for $({\hat g}_{\mu\nu},\psi)$;
\begin{align}
{\hat I}=&\int d^nx\sqrt{-{{\hat g}}}\left[\frac{1}{16\pi G_n}{{\hat R}}-\frac12 ({\hat {\nabla}}\psi)^2-{\hat V}(\psi)\right].
\end{align}
The field equations are
\begin{align}
{\hat G}^\mu_{~\nu}=&8\pi G_n\biggl[{\hat\nabla}^\mu\psi{\hat\nabla}_\nu\psi- \delta^\mu_{~\nu}\biggl(\frac12({\hat\nabla}\psi)^2+{\hat V}(\psi)\biggl)\biggl],\\
{\hat\nabla}^\mu{\hat\nabla}_\mu\psi=&\frac{\partial{\hat V}}{\partial\psi}.\label{psi-ein}
\end{align}
For the following Lifshitz metric with arbitrary $k$ with $\psi=\psi(r)$;
\begin{equation}
{\hat g}_{\mu\nu}dx^\mu dx^\nu = - \frac{r^{2z}}{l^{2z}}  dt^2 + \frac{l^{2}}{r^{2}}  dr^2 + r^2 \gamma_{ij}dz^i dz^j,    \label{L2}
\end{equation}
the field equations becomes
\begin{align}
\frac{(n-2)[(n-1)r^2-(n-3)kl^2]}{16\pi G_nl^2r^2}=&-\frac{r^{2}}{2l^{2}}\biggl(\frac{d\psi}{dr}\biggl)^2-{\hat V}(\psi),\label{ein1}\\
\frac{(n-2)[(n-3+2z)r^2-(n-3)kl^2]}{16\pi G_nl^2r^2}=&\frac{r^{2}}{2l^{2}}\biggl(\frac{d\psi}{dr}\biggl)^2-{\hat V}(\psi),\label{ein2}\\
\frac{[2z(z+n-3)+(n-2)(n-3)]r^2-(n-3)(n-4)kl^2}{16\pi G_nl^2r^2}=&-\frac{r^{2}}{2l^{2}}\biggl(\frac{d\psi}{dr}\biggl)^2-{\hat V}(\psi).\label{ein3}
\end{align}
Equation (\ref{ein1}) minus Eq.~(\ref{ein3}) gives
\begin{align}
(z-1)(z+n-2)r^2+(n-3)kl^2=0.
\end{align}
Hence, $(n-3)k=0$ and $(z-1)(z+n-2)=0$ must be satisfied.
Since $z=1$ gives AdS, we choose $z=2-n$.
Then, Eq.~(\ref{ein1}) plus Eq.~(\ref{ein2}) with $(n-3)k=0$ gives ${\hat V}(\psi)=0$.
With ${\hat V}(\psi)=0$, $(n-3)k=0$, and $z=2-n$, Eq.~(\ref{psi-ein}) reduces to
\begin{equation}
\frac{d}{dr}\biggl(r\frac{d\psi}{dr}\biggl) =0,
\end{equation}
which is integrated to give
\begin{equation}
\psi=\psi_0\ln|r/r_0|,\label{psi-sol}
\end{equation}
where $\psi_0$ and $r_0$ are constants.
On the other hand, Eqs.~(\ref{ein1})--(\ref{ein3}) all give
\begin{align}
\frac{(n-2)(n-1)}{8\pi G_n}=-r^{2}\biggl(\frac{d\psi}{dr}\biggl)^2.
\end{align}
Then, we otain
\begin{align}
\psi_0^2=-\frac{(n-2)(n-1)}{8\pi G_n}.
\end{align}
In summary, in the Einstein frame, the Lifshitz spacetime is not compatible with a real scalar field.
However, the Lifshitz spacetime with $z=2-n$ is compatible with a massless ghost scalar field given by
\begin{equation}
\psi=\pm i\sqrt{\frac{(n-2)(n-1)}{8\pi G_n}}\ln|r/r_0|,\label{psi-sol}
\end{equation}
where $i^2=-1$.

{Let us now turn to the case with non-zero $k$.
In the Einstein-Klein-Gordon system (${\hat V}\equiv 0$), there is an exact static spherically symmetric solution which was originally obtained by Fisher in four dimensions~\cite{fisher}.
This solution was rediscovered by many authors and is often refered as the Janis-Newman-Winicour solution~\cite{jnw}.
(See~\cite{as2010} for the comprehensive study of the arbitrary-dimensional Fisher solution which was given by Xanthopoulos and Zannias~\cite{xz1989}.)
The generalized Fisher-Xanthopoulos-Zannias solution for arbitrary $k$ is given as
\begin{align}
{\hat g}_{\mu\nu}dx^\mu dx^\nu=&-h(\rho)^{a+1}d{t}^2+h(\rho)^{[-a/(n-3)]-1}d\rho^2+\rho^2h(\rho)^{-a/(n-3)}\gamma_{ij}dz^idz^j,\\
\psi=&\pm\sqrt{\frac{-a(a+2)(n-2)}{32\pi G_n(n-3)}}\ln |h(\rho)|,\\
h(\rho)=&k-\frac{m}{\rho^{n-3}},
\end{align}
where $a$ is a parameter and $n\ge 4$.
Actually, the metric is the same as the generalized Brans solution (\ref{brans}) with $b=-a/(n-3)$.
Therefore, from the analysis in the previous section, it is found that this solution is asymptotically Lifshitz for $a=1-n$ and the dynamical exponent is $z=2-n$.
However in this case, the scalar field $\psi$ becomes pure imaginary, namely a ghost scalar field, and $h(\rho)=0$ for $k\ne 0$ is a curvature singularity.
In summary, as the Brans solution, this solution cannot represent an asymptotically Lifshitz black hole.

}

\section{Solution with $z=1$}
\label{app:z=1}
{In this appendix, we present the solution in the special case of $z=1$, in which the background solution is AdS.
The exact solution with $z=1$ is given for $\omega=-n/(n-1)$ as
\begin{align}
ds^2=&-\frac{r^{2}}{l^{2}}f(r)dt^2+\frac{l^2}{r^2}f(r)^{-1}dr^2+r^2dx^idx_i,\\
f(r)=&{\bar k}-M\ln|r|+\frac{64\pi l^{2}Q_{\rm e}^2}{\phi_0\zeta^2r^{n-3}}+\frac{64\pi l^2Q_{\rm m}^2}{\phi_0\zeta^2(n-5)^2r^{5-n}},\label{z=1}\\
\phi=&\frac{\phi_0}{r^{n-1}},\label{scalar-z1}
\end{align}
where $\phi_0$, $l$, ${\bar k}$, and $M$ are constants.
This is the solution with the potential $V(\phi)=V_{\rm M}+V_{\rm Q_{\rm e}}+V_{\rm Q_{\rm m}}$, where 
\begin{align}
V_{\rm M}=&V_0\phi,\quad V_{\rm Q_{\rm e}}=V_1\biggl(\frac{\phi}{\phi_0}\biggl)^{2(n-2)/(n-1)},\quad V_{\rm Q_{\rm m}}=V_2\biggl(\frac{\phi}{\phi_0}\biggl)^{4/(n-1)}.
\end{align}
The constants $V_0$, $V_1$, and $V_2$ are given by 
\begin{align}
V_0=&\frac{M}{l^2},\label{V0-2} \\
V_1=&-\frac{32\pi(n-1)(n-3)Q_{\rm e}^2}{\zeta^2},\label{V1-2} \\
V_2=&-\frac{16\pi(n-1)(n-6)Q_{\rm m}^2}{(n-5)\zeta^2}.\label{V2-2}
\end{align}
In this case, different from the case with $z \ne 1$, the values not only of $Q_{\rm m}$ but also of $Q_{\rm e}^2$ are fixed in terms of the constants in the theory.
For $\phi_0>0$, this spacetime does not represent an asymptotically AdS black hole.
}

\end{document}